\documentclass[11pt,twoside]{article}
\usepackage{./asp2014}

\aspSuppressVolSlug
\resetcounters

\begin{document}

\title {Converged Close-Coupling R-Matrix calculations of Photoionization of Fe XVII in Astrophysical Plasmas: from Convergence to Completeness}

\author {L. Zhao, $^1$ W. Eissner, $^2$ S. Nahar, $^1$ and A. Pradhan $^1$ \\
\affil {$^1$The Ohio State University, Columbus, Ohio, USA; \email {zhao.1157@osu.edu}; \email{nahar.1@osu.edu}; \email{pradhan.1@osu.edu}}
\affil{$^2$1. Institut f\"ur Theoretische Physik, Universit\"at Stuttgart, Stuttgart, Baden-W\"urttemberg, Germany; \email {we@theo1.physik.uni-stuttgart.de}}}

\paperauthor{L. Zhao}{zhao.1157@osu.edu}{}{The Ohio State University}{Department of Physics}{Columbus}{Ohio}{43210}{USA}
\paperauthor{W. Eissner}{we@theo1.physik.uni-stuttgart.de}{}{Universit\"at Stuttgart}{Institut f\"ur Theoretische Physik}{Stuttgart}{Baden-W\"urttemberg}{70550}{Germany}
\paperauthor{S. Nahar}{nahar.1@osu.edu}{}{The Ohio State University}{Department of Astronomy}{Columbus}{Ohio}{43210}{USA}
\paperauthor{A. PRADHAN}{pradhan.1@osu.edu}{}{The Ohio State University}{Department of Astronomy}{Columbus}{Ohio}{43210}{USA}

\begin{abstract}
Extensive resonance structures are manifest in R-Matrix (RM)
calculations. However, there exist a large number of highly excited electronic
configurations that may contribute to background non-resonant 
bound-free opacity in high-temperature plasmas. Since RM calculations are 
very complex, and not essential for background contributions, the
Relativistic Distorted Wave (RDW) method is utilized to complement
("top-up") photoionization cross sections of Fe XVII obtained using
Close-Coupling Breit-Pauli R-Matrix (CC-BPRM) method. 
There is good agreement between
RDW and BPRM for
background cross sections where resonances are not present,
and individual fine structure levels
can be correctly matched spectroscopically, 
though resonances are neglected in the RDW. 
To ensure completeness, a high energy range up to 500 Ry 
above the ionization threshold for each level is considered.
Interestingly, the hydrogenic Kramer's approximation also shows the same
energy behavior as the RDW. Grouping separately,
the BPRM configurations consist of 454 bound levels with resonances
corresponding to configurations $1s^22s^22p^4nln'l'$ (n $\leq$ 3, n'
$\leq$ 10); including RDW
configurations there are 51,558 levels in total. The topup contribution results in $\sim$20\% increment, in addition to 
the 35\% enhancement from BPRM calculations over the Opacity Project 
value for the Rosseland Mean Opacity at the Z-temperature of 
2.11 $\times 10^6$K \citep{pn_2017}..

\end {abstract}

\section {Introduction}
Iron opacity at the condition similar to the solar radiation/convection zone boundary was measured in Sandia National lab \citep {sandia_2015}, 
revealing that the iron opacity is up to 30-400\% higher than that predicted
by theoretical opacity  models. 
To resolve this discrepancy, extensive 
Close-Coupling R-Matrix calculations with 60 fine structure levels of
the core ion Fe~XVIII with $n \leq 3$ \citep {60cc_2011}, and 99 LS terms with $n \leq 4$
were carried out, showing strong photon absorption due to 
core excitation and resulting in an increment of 35\% in the
Rosseland mean opacity over the Opacity Project (OP) data
\citep [] [hereafter NP16]{99cc_2016}. 
Whereas the NP16 work demonstrated that in R-Matrix opacity calculations 
convergence of the CC wavefunction expansion 
is a necessary condition for accuracy, sufficiency in terms
of completeness of all possible excited configurations at Z plasma
temperatures still requires additional contrbutions in the high-energy
range \citep{comment_2016, comment_2016a, more_comment_2017}. 

In this paper, we address completeness issue as follows. We consider the
60-level BPRM calculations since (i) fine structure is included and is
important, and (ii) the 99LS calculations show background 
convergence with additional resonances converging on 
to the $n = 4$ Fe~XVIII levels,  but do not result in
further enhancment of the Fe~XVII Rosseland Mean Opacity, most
likely owing to neglect of fine structure.
The 60 CC BPRM \citep {60cc_2011} levels have also been 
identified (accessible through {\footnotesize {NORAD Atomic Data (Fe XVII)} \url {http://norad.astronomy.ohio-state.edu/fe17/fe17.en.fs.txt}}). 
In the following sections, the 60CC data
included in \citet {60cc_2011} are displayed, followed by the 
RDW top-up configurations and transitions 
calculated using the flexible atomic code (FAC), an open-source software 
package \citep {gu_2008}. To ensure the correspondence of 
the data from FAC, comparison of the photoionization cross section from both 
FAC and BPRM is investigated. Moreover, to also ensure completeness in
energy, calculation in the high energy region is extended to 500 Ry using
FAC.

\section {Configurations in 60 CC BPRM}
The 60 CC BPRM calculation of photoionization of Fe XVII is
comprehensively studied in \citet {60cc_2011}, and the 454 bound levels
as well as the 60 target states are readily available in {\footnotesize
{NORAD Atomic Data (Fe XVII)} \url
{http://norad.astronomy.ohio-state.edu/fe17/fe17.en.fs.txt}}. The bound
and target configurations are in the below (the full K-shell is omitted for simplicity):
\begin {itemize}
\item Bound Configurations:
	\begin {itemize}
	\item $2s^2 2p^6$
	\item $2s^2 2p^5$ with $3l,~4l,~5l,~6l,~7l,~8l,~9(s-k),~10(s-k)$, where $l$ represents all the subshells in a shell
	\item $2s 2p^6$ with $3l,~4l,~5s$
	\end {itemize}
\item Target Configurations:
	\begin {itemize}
	\item $2s^2 2p^5$
	\item $2s 2p^6$
	\item $2s^22p^4 3l$
	\end {itemize}
\end {itemize}

In the next section, all the possible top-up transitions do not involve any one appearing in 60 CC BPRM.

\section {Top-up Configurations and Transitions }
To complement the photoionization cross section of 60 CC BPRM, 
part of the top-up configurations and transitions are taken from \citet {config_2003}, \citet {config_2005} and \citet {more_comment_2017}. Here are the top-up configurations and transitions in the following below:
\begin {itemize}
\item 454-level L-shell photoionization
	\begin {itemize}
	\item The same 454 levels as in 60 CC BPRM, but are considered as L-shell photoionization only, excluding any transition that appears in 60 CC BPRM.
	\end {itemize}
\item The other L- and outer-shell photoionization
	\begin {itemize}
	\item $2s 2p^6$ with $5(p-g),~6l,~7l,~8l,~9l~\text{and}~10l$.
	\item One type of transition is L-shell photoionization, which ionizes one electron in L-shell, leaving the rest intact.
	\item The other type of transitions is "outer-shell" photoionization, with any dipole-allowed final states from $2s2p^6,~2s^22p^5,~2s^22p^43l$.
	\end {itemize}
\item 2-hole Configurations \\
 Each 2-hole configuration can be divided into two parts, inner part and outer part, and the 2-hole configurations included are formed by combining any one in inner part and any one in outer part.
	\begin {itemize}
	\item Inner Part:
		\begin {itemize}
		\item $2p^6$
		\item $2s2p^5$
		\item $2s^22p^4$
		\end {itemize}
	\item Outer Part:
		\begin {itemize}
		\item $3l^2$, $3l4l'$, $3l5l'$, $3l6l'$
		\item $4l^2$, $4l5l'$, $4l6l'$
		\item $5l^2$, $5l6l'$
		\item $6l^2$
		\end {itemize}
	\end {itemize}
\end {itemize}

The transitions considered from the 2-hole configurations are: i) in the outer \indent part either of the two electrons is ionized, leaving the inner part intact; ii) one \indent electron is ionized from the L-shell. \\\\
Using FAC, it's fairly easy to implement the photoionization
calculations compared with the BPRM, resulting in 51,558 initial levels
and more pairs of transitions among them. 
For each level, an individual energy mesh is created according to the 
ionization thresholds so that the photoionization cross section is well 
resolved at all energies. The final photoionization cross section 
for a level is the sum of all the transitions from the level in the same
energy mesh, up to 500 Ry above the lowest threshold.

\section {Comparison between RDW and BPRM}
To top up the photoionization cross section for 454 levels included in
60 CC BPRM \citep {60cc_2011} by including additional 
transitions from the L-shell, energy levels have to be matched between 
RDW and BPRM. By comparing J, Pi, energy order and some other factors, they are readily matched. To ensure their being matched correctly, RDW calculation is done on the photoionization cross section from the 454 levels to any possible core states included in 60 CC BPRM, i.e. levels in $2s^2 2p^5$, $2s 2p^6$ and $2s^22p^4 3l$. Generally the RDW matches well with the background of the BPRM result, 
but misses out all the resonances in BPRM (see Figure \ref {bprm_rdw}). \\

\articlefigure [width=.8\textwidth] {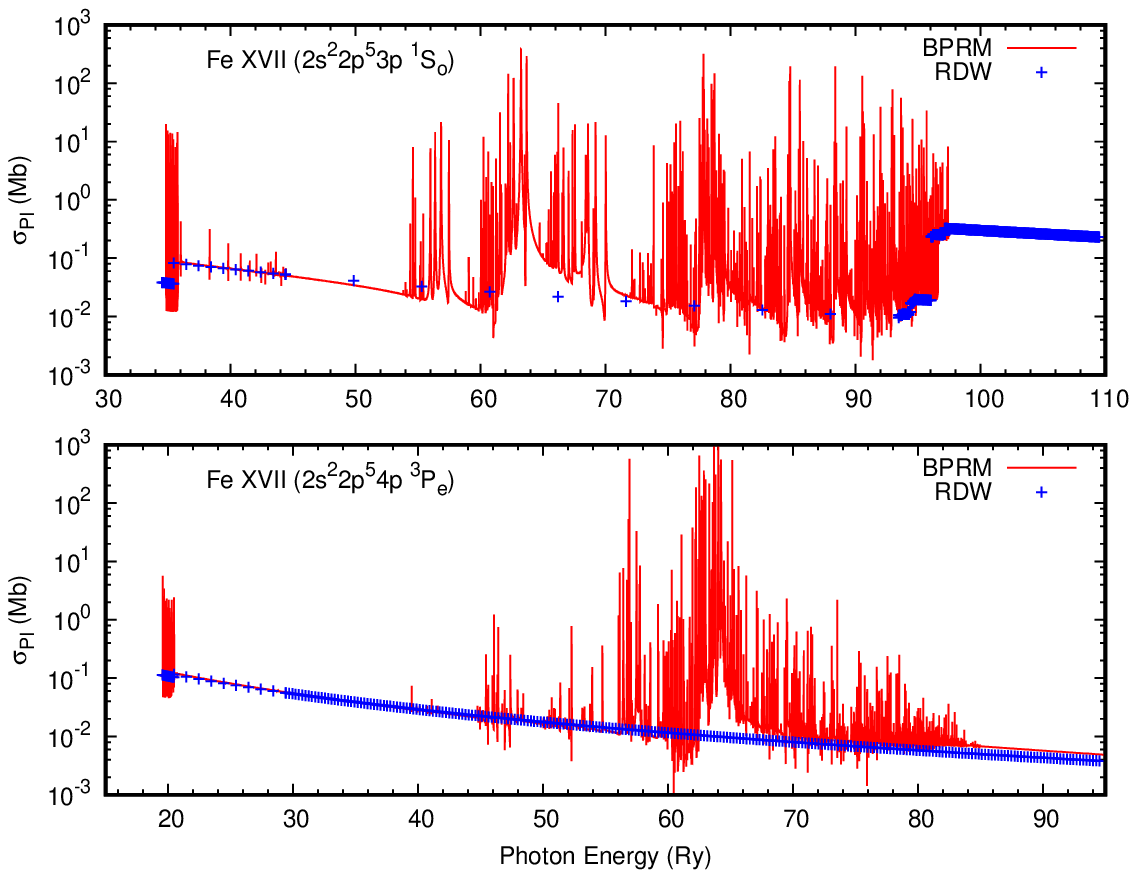} {bprm_rdw} {Comparison of the photoionization cross section for two levels using both RDW and BPRM. \emph {Above:} Fe XVII $2s^22p^53p~^1S_o$. \emph {Below:} Fe XVII $2s^22p^54p~^3S_e$}

\articlefigure [width=.8\textwidth] {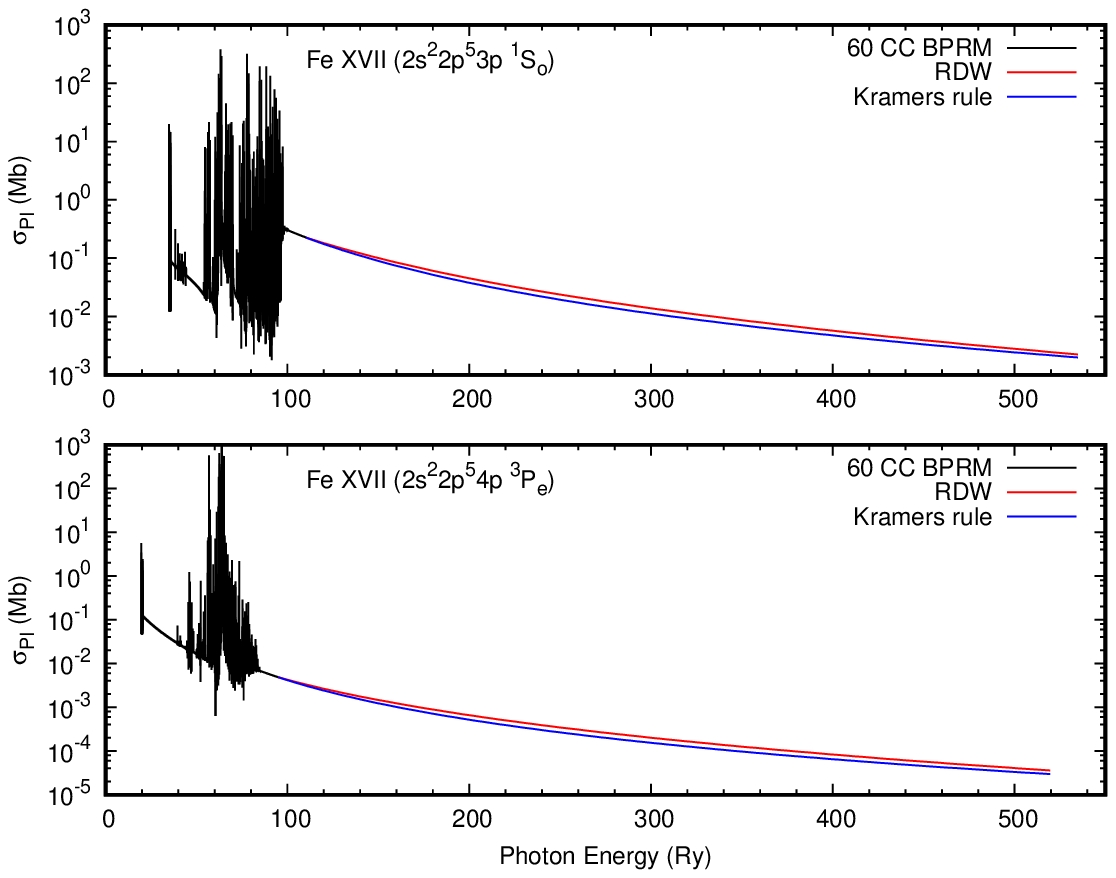} {rm_fac_kramer} {Photoionization cross section in high energy regime using RDW and Kramer's rule. \emph {Above:} Fe XVII $2s^22p^53p~^1S_o$. \emph {Below:} Fe XVII $2s^22p^54p~^3S_e$}

\section {Photoionization Cross section in high energy region}
In addition to adding more possible configurations and transitions,
energy range in which the calculation is done is extended to 500 Ry 
above the lowest energy threshold for each level. As the BPRM computation in 
such a large energy range is expensive, and also unncecessary,
the  RDW is used to extend the high-energy 
"tails" of photoionization cross section for the 454 levels in 60 CC BPRM. 
The RDW data are rescaled by the ratio of photoionization cross section from BPRM and RDW at the last energy point in BPRM, as the BPRM data should be more accurate. 
Kramer's rule ($\sigma_\nu =
\sigma_1(\nu_1/\nu)^3$) is also applied to extrapolate in
the tails region. It turns out that
the opacity contribution from the high energy regime is equivalent using RDW and Kramer's rule (see Figure \ref{rm_fac_kramer}). \\

\section{Conclusion}
Calculation of photoionization cross section of Fe XVII is completed 
by including more top-up configurations and transitions, and extending 
into the high-energy bound-free continuum. We find an additional
$\sim$20\% enhancement, in addition to the 35\% reported in NP16, with the total topped-up result of 1.64 times the 
OP value for the Rosseland mean opacity at the Z temperature \citep {pn_2017}. The
high-energy 
topup is thus close to the 16\% estimated in \citep{more_comment_2017}. 
However, the actual Fe~XVII Rosseland mean opacity might be still larger due to
additional fine structure thresholds from the 218-level BPRM calculation in
progress including resonances converging on to the $n = 4$ thresholds of
Fe~XVIII.
\acknowledgements
This research was supported by a teaching assistantship from the
Department of Physics and the Department of Astronomy in the Ohio State
University (OSU), and the U.S. National Science Foundation and the Department
of Energy. The computations were carried out 
at the Ohio Supercomputer Center (OSC) in Columbus Ohio, and the
OSU Department of Astronomy.


\begin {thebibliography} {}
\bibitem [Badnell  et al. (2005)] {config_2005} Badnell, N.R., Bautista,
M. A., Butler, K., et al. 2005, MNRAS, 360, 458 
\bibitem [Badnell \& Seaton (2003)] {config_2003} Badnell, N. R., \& Seaton, M. J. 2003, J. Phys. B, 36, 4367
\bibitem [Bailey et al. (2015)] {sandia_2015} Bailey, J., Nagayama, T., Loisel, G. P., et al. 2015, Nature, 517, 56
\bibitem [Blancard et al. (2016)] {comment_2016} Blancard, C., Colgan, J., Coss\'e, Ph., Faussurier, G., et al. 2016, PRL, 117, 249501
\bibitem [Gu (2008)]{gu_2008} Gu, M. F. 2008, Can. J. Phys., 86, 675
\bibitem [Iglesias \& Hansen (2017)] {more_comment_2017} Iglesias, C., \& Hansen, S. 2017, ApJ, 835, 5
\bibitem [Nahar, et al. (2011)] {60cc_2011} Nahar, S. N., Pradhan, A. K., Chen, G. X., \& Eissner, W. 2011, Physical Review A, 83, 053417
\bibitem [Nahar \& Pradhan (2016)] {99cc_2016} Nahar, S. N., \& Pradhan, A. K. 2016, PRL, 116, 235003
\bibitem [Nahar \& Pradhan (2016)] {comment_2016a} Nahar, S. N., \& Pradhan, A. K. 2016, PRL, 117, 249502
\bibitem[Pradhan \& Nahar (2017)]{pn_2017} Pradhan, A., \& Nahar, S. In the same volume. \\


\end {thebibliography}
\end {document}